\documentclass[12pt]{article}
\usepackage{pic04}
\usepackage{hyperref}
\usepackage{url}
\usepackage{graphicx}

\begin{document}

\title{\bf TEVATRON ELECTROWEAK RESULTS AND \\ ELECTROWEAK SUMMARY}
\author{
Sean E. K. Mattingly\\
{\em Brown University, Physics Department, 182 Hope St., Providence, RI 02912}\\
For the CDF and DZero Collaborations}
\maketitle

%
%
%
%
%
%
\vspace{4.5cm}
%

\baselineskip=14.5pt
\begin{abstract}
Electroweak measurements from early in Run II of the Tevatron Collider at Fermilab are presented as well as future
prospects for Run II. 
The effect of Tevatron measurements on worldwide electroweak fits and the Higgs mass is also discussed.
\end{abstract}
\newpage

\baselineskip=17pt

\section{Introduction}

Electroweak measurements by the CDF and DZero collaborations on the Tevatron at Fermilab center around $W$ and $Z$ 
boson production in proton-antiproton collisions. Only the leptonic decays of the bosons are considered since
these channels have relatively small backgrounds. The well known $W$ and $Z$ boson production cross sections are high, 
and when they are combined with the clean
decay signatures, precision electroweak measurements are attainable.

Precision measurements of the $W$ and $Z$ boson cross sections times branching ratios
in the leptonic decay channels not only test the
Standard Model description of the production process, but they can also be used to test lepton universality by comparison
of the different channels.
The ratio of the $W$ and $Z$ production cross sections is used as an indirect
measurement of the $W$ boson mass and width. 
With sufficiently precise measurements of $W$ boson production, the transverse mass distribution can be used as a direct
measurement of the mass and width of the $W$ boson and the charge asymmetry of the produced $W$ bosons constrains
the proton parton distribution functions. In $Z$ boson production, the forward backward asymmetry yields a 
measurement of $sin^2(\theta_W)$ and the $Z$-quark couplings, and searches for higher mass $Z'$ bosons are
conducted. Finally, measurements of diboson production processes 
({\em i.e.} $WW$, $WZ$, $W\gamma(\gamma)$ and $Z\gamma(\gamma)$ production) are used to study triple and quartic
gauge couplings. Production of $W$ and $Z$ bosons is an important background to studies of the top quark
and searches for the Higgs boson and SUSY particles, where detailed understanding of the backgrounds is important.

The upgrades to the CDF and DZero detectors for Run II, combined with \mbox{$\sim400$ pb$^{-1}$} of luminosity delivered by the
Tevatron by summer 2004, with instantaneous luminosties above $\sim6 \times 10^{31}/$cm$^2$s and rising, give 
the collaborations the ability to conduct electroweak measurements at higher precision than ever before available
at a hadron collider. The results presented here represent the status of electroweak analyses at CDF and DZero
as of June 2004 and use up to 200 pb$^{-1}$ of the initial luminosity delivered by the Tevatron.

\section{$W$ and $Z$ Boson Cross Sections}

Decays of the $Z$ boson are identified by observing two high $p_T$ leptons with an inviariant mass 
near the $Z$ boson
mass. 
Decays of $W$ bosons are distinguished by one high $p_T$ lepton and
missing transverse energy from the 
escaping neutrino. Since the longitudinal momentum of the neutrino cannot be unambiguously determined, the
lepton-neutrino invariant mass cannot be calculated. However, since the transverse momentum of the neutrino is
estimated using the missing transverse energy, the transverse mass, 
$M_T = \sqrt{ 2E_T^\ell E_T^{missing} \left[ 1 - \cos(\Delta\phi^{\ell -missing}) \right] }$, can be calculated 
and is expected to be high.

The electronic and muonic decay channels primarily used for these analyses offer the cleanest
signatures. Electronic decays are triggered by a deposition of energy in the electromagnetic layers of the
calorimeter. Final electron identification requires the deposition of energy to be consistent with an
electromagnetic shower and generally requires a track to be matched to the energy cluster in the calorimeter.
Muonic triggers require a muon to be found in the muon chambers and a track with high $p_T$ to be seen in the inner tracker. 
Muons used in analyses are required to have an isolated high $p_T$ track matched either to a muon
detector track or to the track of a muon as a minimum ionizing particle passing through the calorimeter. The
efficiencies of the triggers and object identification are measured using $Z \rightarrow \ell^+\ell^-$ events
where one lepton has passed all the identification and trigger requirements, leaving the other lepton as an
unbiased probe for studying efficiencies. The primary backgrounds to $W$ and $Z$ boson production come from
QCD jets and the background contributions are estimated using jet data.

The cross section times branching ratio for $Z \rightarrow e^+e^-$ is measured by both experiments
using electrons with greater than 25 GeV of transverse energy. DZero uses electrons with no track
match required and $|\eta| < 1.1$ and the invariant mass of the electron pair is required to be between 60 and
110 GeV.
The measurement by CDF counts events in a mass window between 65 and 115 GeV and 
requires tracks to be matched to both electrons, with one electron required to be central while
the other electron is allowed to be forward. There are
small backgrounds from jets faking electrons and from $Z$ boson decays to $\tau^+\tau^-$ where the taus decay
electronically. DZero uses 41 pb$^{-1}$ of luminosity while CDF uses 71 pb$^{-1}$, yielding 1139 and 4242 events
respectively. 
The cross section times branching ratio calculated by CDF is 
$255.2 \pm 3.9(stat) 
\pm ^{5.5}_{5.4}(syst)
\pm 15.3(lumi)$ pb.
The measurement reported by DZero includes a small correction for the contribution from Drell-Yan
production and is found to be
$275 \pm 9(stat) 
\pm 9(syst)
\pm 28(lumi)$ pb.

For the $Z \rightarrow \mu^+\mu^-$ cross section times branching ratio measurement, CDF uses muons above
20 GeV with $|\eta| <$ 1.0 while DZero uses muons above 15 GeV with $|\eta| <$ 1.8. This channel has very
small backgrounds from QCD b-quark jets that decay semi-leptonically to muons, cosmic muons, 
and $Z \rightarrow \tau^+\tau^-$ decays where the taus decay muonically. CDF uses 72 pb$^{-1}$ of
data and finds 1371 events in a 65 to 115 GeV invariant mass window, while DZero uses 117 pb$^{-1}$ of data
and finds a total of 6126 events. 
The CDF measurement of the cross sesction times branching ratio is
$248.9 \pm 5.9(stat) 
\pm ^{7.0}_{6.2}(syst)
\pm 14.9(lumi)$ pb
and the DZero result is
$261.8 \pm 5.0(stat) 
\pm 8.9(syst)
\pm 26.2(lumi)$ pb.
These cross sections are in good agreement with the measurements of $Z$ boson production in the electron channel.

Measurements of the $W$ boson production cross sections are also made in the electron and muon decay channels. For the measurement in the
electron channel, both CDF and DZero look for an electron with above 25 GeV of transverse energy in events with above 25 GeV of missing transverse
energy. In 42 pb$^{-1}$ of data, DZero requires the electron to be central and selects $\sim$27,400 events, 
while CDF uses central and forward electrons in two separate analyses
that use 72 pb$^{-1}$ of data and finds a total of $\sim$48,000 events. 
The main background comes from jet 
events with a jet faking an electron and mismeasurement leading to missing transverse energy. Secondary backgrounds come from $W \rightarrow \tau\nu$
decays where the tau decays electronically and from $Z \rightarrow e^+e^-$ decays where one electron is missed. DZero measures
a cross section times branching ratio of $2844 \pm 21(stat) \pm 128(syst) \pm 284(lumi)$. Using central
electrons, CDF measures $2782 \pm 14 \pm ^{61}_{56}(syst) \pm 167(lumi)$ and using forward electrons finds 
$2874 \pm 34(stat) \pm 167(syst) \pm 172(lumi)$.

Measurements for the cross section times branching ratio for $W \rightarrow \mu\nu$ use events with missing transverse
energy greater than 20 GeV and muons having $p_T > 20 GeV$. DZero uses muons with $|\eta| < 1.6$ while CDF
uses muons with $|\eta| < 1.0$. The DZero analysis uses the initial 17 pb$^{-1}$ of Tevatron luminosity and observes $\sim$8,300 events 
while the CDF result uses 72 pb$^{-1}$ of luminosity and finds $\sim$31,700 events. The resulting cross sections times
branching ratio are 
$3226 \pm 128(stat) \pm 100(syst) \pm 322(lumi)$ from DZero and
$2772 \pm 16(stat) \pm ^{64}_{60}(syst) \pm 166(lumi)$ for CDF. As with the measurements for $Z$ boson production, the measurements
for $W$ boson production in the electron and muon channels are in good agreement.

\begin{figure}[htbp]
  \centerline{\hbox{ \hspace{0.2cm}
    \includegraphics[width=6.5cm]{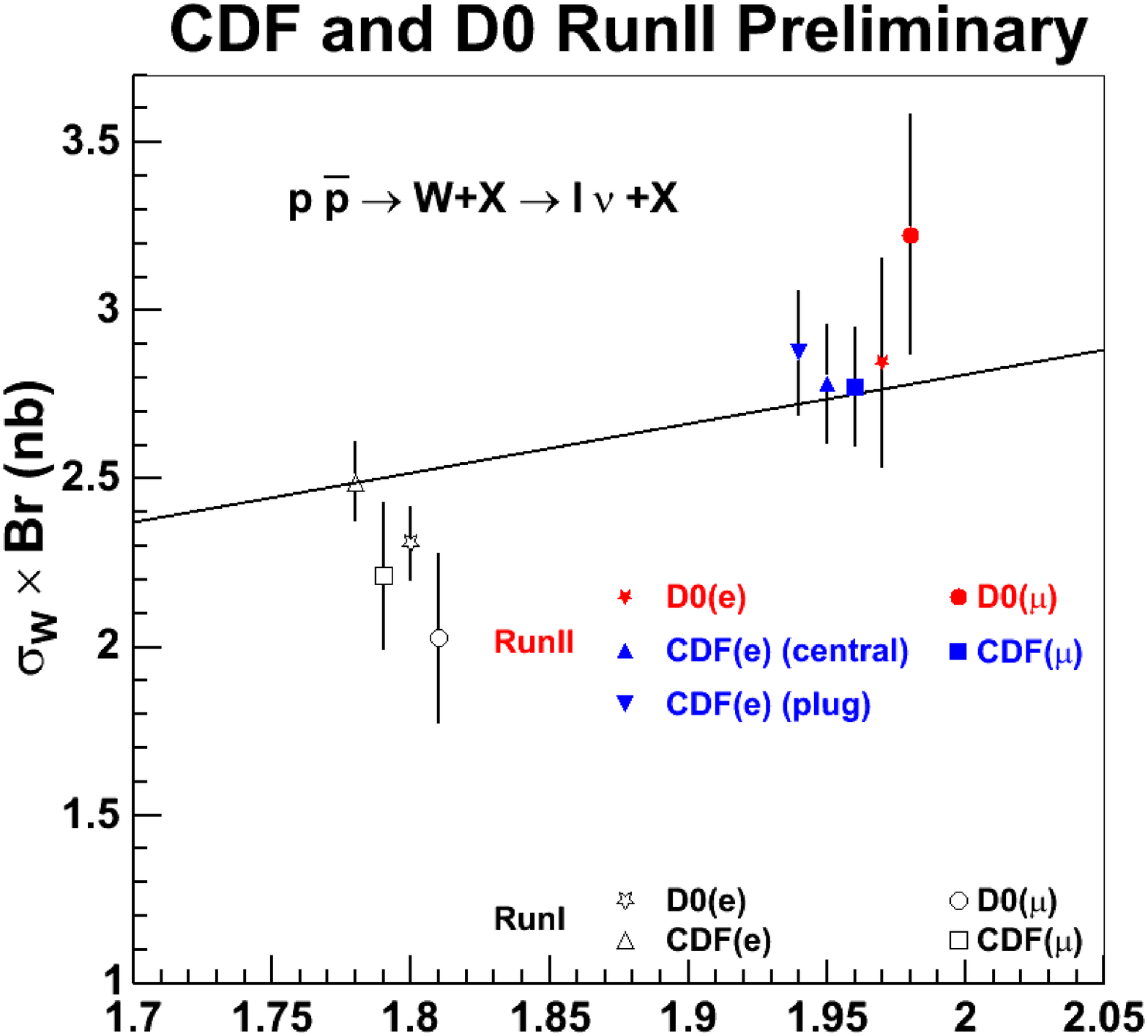}
    \hspace{0.3cm}
    \includegraphics[width=6.5cm]{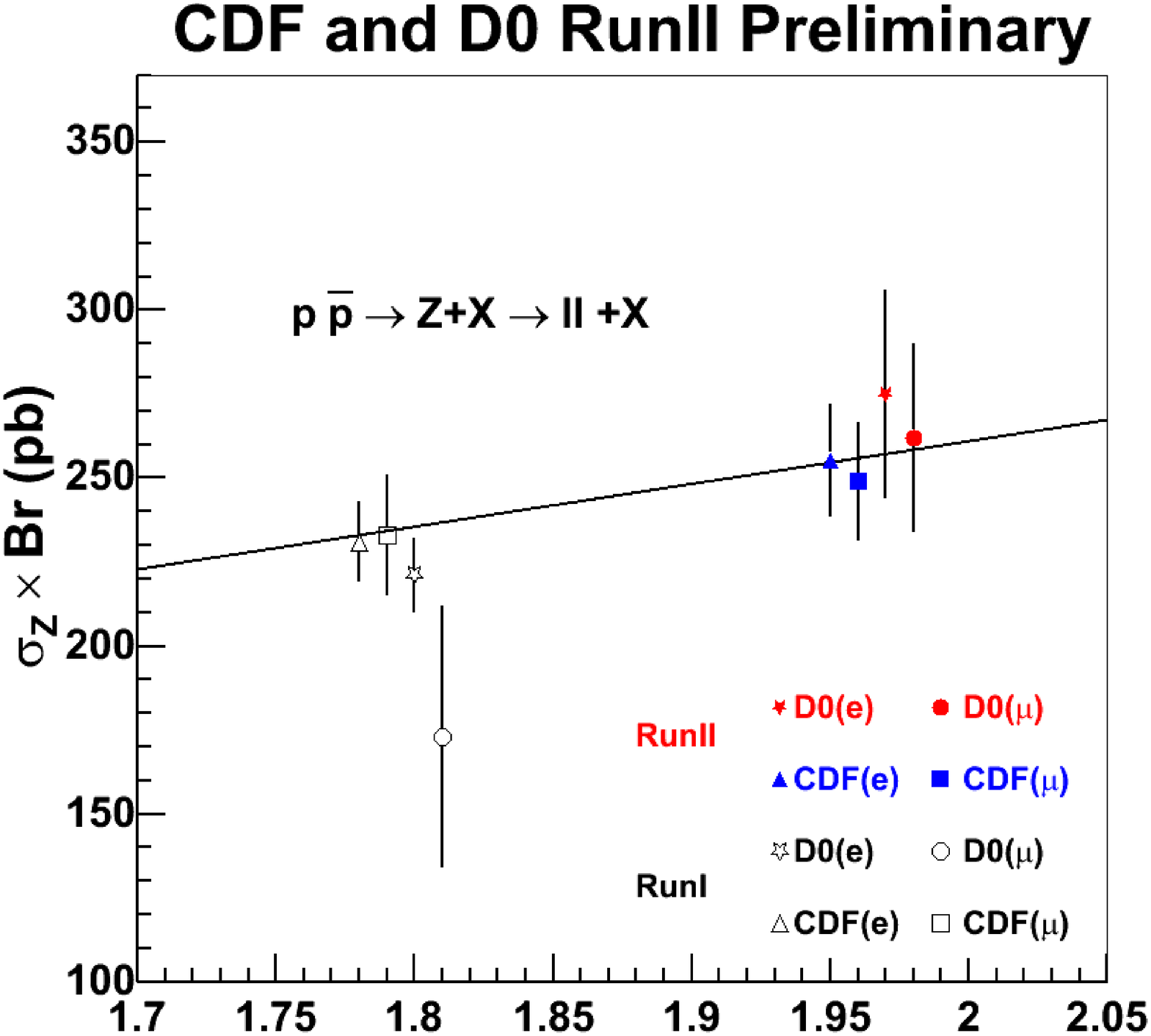}
    }
  }
 \caption{\it
      Measurements of $W$ and $Z$ boson cross sections at the Tevatron as a function of center of mass energy.  The lines are NLO predictions
by Van Neerven and Matsuura.
    \label{wzxsecs} }
\end{figure}

The efficiencies and purities attained by CDF and DZero are similar except where CDF includes forward electrons and where DZero uses
forward muons. These differences will soon be resolved as CDF will include farther forward muons and DZero will include forward electrons.
A summary of the Run II cross sections is shown 
in figure \ref{wzxsecs} along with the cross sections measured in Run I and NLO predictions by Van Neerven and Matsuura. Both the Run I and Run II 
measurements are in agreement with the theoretical predictions.

By using the measured ratio of $W$ and $Z$ boson cross sections times branching ratios and theoretical inputs, an indirect measurement of the
$W$ boson branching ratio to leptons and the full $W$ boson width can be made using the relation
\begin{equation}
R \equiv \frac{\sigma_W \times BR(\ell\nu)}{\sigma_Z \times BR(\ell\ell)} 
= 
\frac{\sigma_W}{\sigma_Z}\frac{\Gamma(Z)}{\Gamma(Z \rightarrow \ell\ell)}
\frac{\Gamma(W \rightarrow \ell\nu)}{\Gamma(W)}.
\end{equation}
The values of $\sigma_W$ and $\sigma_Z$, the total $W$ and $Z$ boson production cross sections, are obtained from an NNLO calculation by Van Neerven,
the $Z$ boson branching ratio $\Gamma(Z)/\Gamma(Z \rightarrow \ell\ell)$ is taken from LEP measurements and the partial $W$ boson width is
a Standard Model electroweak calculation. Following this procedure after combining the CDF electron and muon channels yields
$R = 10.93 \pm 0.15(stat) \pm 0.13(syst)$,
$BR(W \rightarrow \ell\nu) = (10.93 \pm 0.21)\%$ and
$\Gamma(W) = 2.071 \pm 0.040$ GeV, which are in agreement with the world average values.

As a check of lepton universality, the $W$ and $Z$ boson cross sections are measured in the tau decay channels.  Both CDF and
DZero measure the cross section times branching ratio for $Z \rightarrow \tau^+\tau^-$ using events where one tau decays
leptonically and the other tau follows a one-prong hadronic decay.  DZero looks for the leptonic decay of a tau in the muon
channel in 68 pb$^{-1}$ of data and finds a cross section times branching ratio of $222 \pm 36(stat) \pm 57(syst) \pm 22(lumi)$ pb.
The measurement by CDF uses electronic decays of the tau and a luminosity of 72 pb$^{-1}$ and the result is 
$242 \pm 48(stat) \pm 26(syst) \pm 15(lumi)$ pb.  These measurements are in agreement with the cross sections measured in
the other lepton channels and demonstrate the ability to observe resonances in di-tau final states.
CDF has also measured the cross section times branching ratio for 
$W \rightarrow \tau\nu$ production.  The analysis uses events triggered by events with a central track and missing transverse energy.
Reconstruction of the tau decay is accomplished by finding an energy cluster in the calorimeter with  tracks matched to it
within a 10$^\circ$ cone and no tracks between the 10$^\circ$ cone and a 30$^\circ$ cone.  Individual $\pi^0$ particles from the tau
decay are
reconstructed using detectors in the calorimeter at shower max, and the combined invariant mass of the cluster is required to
be less than the mass of the tau.  After accounting for backgrounds from $W \rightarrow \mu\nu$, $W \rightarrow e\nu$, 
$Z \rightarrow \tau\tau$, and jets, a cross section times branching ratio of $2.62 \pm 0.07(stat) \pm 0.21(syst) \pm 0.16(lumi)$ nb
is measured.  Assuming a consistent $W$ boson production cross section, this measurement implies that the ratio of the
coupling of the $W$ boson to the tau to the coupling to the electron is $0.99 \pm 0.04$.

Higher precision measurements are possible with improvements in a number of areas.  The largest
error in many measurements is the luminosity error which is as large as 10\% in some cases.  The understanding of the luminosity
measurement is improving, and CDF and DZero now use identical luminosity determinations, which have an improved
error of 6.5\%.  With the record luminosities provided by the Tevatron, large data sets are becoming available.  
Collection and analysis of additional data will not only improve the statistical errors but will also lead to smaller systematics
from lepton identification as well as refined estimates of background contributions.  Improvements in the understanding of
detector response and material descriptions will also increase the precision of the measurements.  Understanding
of the proton parton density functions and the associated systematic errors is improving with use of the recently
available full error treatments in the CTEQ6 and MRST parton density function sets.  Combining the results from
CDF and DZero with full accounting for correlated errors will yield the most precise measurements.  To that end, the Tevatron
Electroweak Working Group has been formed and has produced early combined Tevatron results \cite{tevewwg}.

\section{Measurements Using $W$ and $Z$ boson production}

\subsection{Asymmetries}

Measurements of various angular asymmetries in $W$ and $Z$ boson production are sensitive to details of electroweak
physics as well as the parton distribution functions of the proton.  In the process 
$p\overline{p} \rightarrow Z/\gamma^* \rightarrow \ell^+\ell^-$, the asymmetry
\begin{equation}
A_{fb} = \frac{\sigma(\cos\theta > 0) - \sigma(\cos\theta < 0)}{\sigma(\cos\theta > 0) + \sigma(\cos\theta < 0)},
\end{equation}
where theta is the polar angle between the incoming proton and the outgoing lepton, is sensititive to the V-A coupling to the
$Z$ boson.
Measurements of $\sin^2 \theta_W$ and the $u$ and $d$ quark couplings can be made by fitting to the shape of this asymmetry
as a function of center of mass energy.  This analysis has been performed by CDF using electronic decays of the $Z$ boson
and finds $\sin^2\theta_W$ is $0.2238 \pm 0.0046(stat) \pm 0.0020(syst)$ and quark couplings consistent with the
Standard Model values.

\begin{figure}[htb]
\centerline{
\includegraphics[width=7cm]{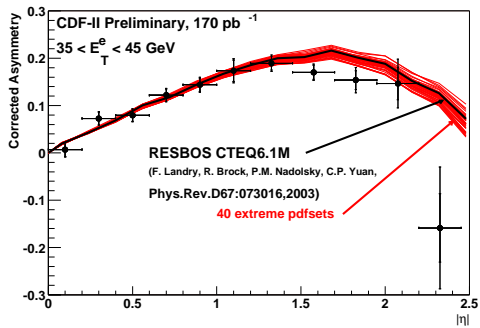}
}
\caption{\it The charge asymmetry as a function of $|\eta|$ for $W \rightarrow e\nu$ as measured by CDF.  The set of curves
represents the uncertainty in CTEQ6 using the 40 error parton distribution functions.}
\label{wasym}
\end{figure}

Since up-type quarks carry more average momentum than down-type quarks, $W^+$ bosons tend to be boosted in the proton direction
while $W^-$ bosons tend to be boosted in the antiproton direction.  Thus, a measurement of the charge asymmetry
\begin{equation}
A_W = \frac{d\sigma(W^+)/dy - d\sigma(W^-)/dy}
           {d\sigma(W^+)/dy - d\sigma(W^-)/dy}
\end{equation}
can be used to constrain the parton distribution functions of the proton, particularly for the $u$ and $d$ quarks at high $x$.  
Since the neutrino in $W \rightarrow e\nu$ decays
escapes the detector unobserved, the direction of the produced $W$ boson cannot be unambiguously determined.  However, the
direction of the outgoing electron/positron is correlated, with increasing correlation with increasing
transverse energy, to the $W$ boson's direction.  So, the asymmetry is measured using the
rapidity of the electron/positron.  The $W$ charge asymmetry has been measured by the CDF collaboration using $W$ boson events
where the transverse mass is between 50 and 100 GeV and no other electromagnetic object with $E_T > 25$ GeV is allowed.  The
charge of the track matched to the electromagnetic calorimeter is used to determine whether the outgoing particle was an electron
or a positron.  Forward silicon detectors are used with calorimeter seeded tracking to determine the charge up to $|\eta| < 2$,
and the charge misidentification rate determined from $Z \rightarrow e^+e^-$ events is $<1\%$ for the central region and
$<4\%$ farther forward.  Backgrounds bias the asymmetry toward zero and are subtracted using Monte Carlo techiniques for
$Z \rightarrow e^+e^-$ and $W \rightarrow \tau\nu$ backgrounds and using the data for jet backgrounds.  Figure \ref{wasym}
shows the
resulting measurement as well as the uncertainty range obtained from the CTEQ6 error parton distribution functions.  As expected,
the largest sensitivity in this measurement is for forward electrons/positrons with high transverse energy.  As this measurement
is refined, strong constraints will improve the uncertainties of the parton distribution functions.

\subsection{Drell-Yan invariant mass spectrum}

The invariant mass spectrum of dilepton pairs in $p\overline{p} \rightarrow Z/\gamma^* \rightarrow \ell^+\ell^-$ is described
well by theoretical predictions at low mass where statistics are high.
Deviations from the predictions would indicate
new physics, and many postulated processes are searched for at high invariant mass.  For example, both DZero and CDF have
searched for a 
$Z'$ boson.  No excesses
are observed and DZero sets a 95\% confidence level limit on the mass of a $Z'$ boson with Standard Model couplings
to be greater than 780 GeV.  A 95\% confidence level limit of 735 GeV is set by the CDF analysis.  Limits on other new
physics are also set using the Drell-Yan invariant mass spectrum and are reported on elsewhere in these proceedings \cite{nagy}.

\section{Diboson Production}

Collisions at the Tevatron can produce multiple bosons.  These events can be used to probe the gauge structure of electroweak
physics and search for anomalous couplings.  Diboson production is an important background in searches for new physics and
improved modeling of diboson production will aid in those searches.  As in measurements of $W$ and $Z$ single boson production,
DZero and CDF focus on the electronic and muonic decays of the bosons in order to minimize backgrounds.

In $W\gamma$ production the photon can come from initial state radiation from one of the incoming quarks,
final state radiation from an outgoing lepton, or radiation directly from the $W$ boson.  This final source
of photons is dependent on the $WW\gamma$ triple gauge coupling and destructively interferes with the initial
state radiation contribution.  If this coupling is different than that in the Standard Model, an excess of photons at high transverse
energy is expected.  Both CDF and DZero measure $W\gamma$ production in the electron and muon channels by first selecting
$W \rightarrow \ell\nu$ events and adding a photon requirement.  Photons are identified as isolated electromagnetic showers in
the central calorimeter with no track matched to them and having separation $\Delta R = \sqrt{\Delta\phi^2 + \Delta\eta^2} > 0.7$
from the outgoing lepton to reduce the influence of final state radiation.  
Photons in the CDF analysis are required to have $E_T > 7$ GeV while DZero
requires $E_T > 8$ GeV.  The primary background is from $W$+jet events where the jet leaves an electromagnetic deposit in the
calorimeter and fakes a photon.  Smaller backgrounds come from $Z+\gamma$ and $Z$+jet events where one lepton from the $Z$
boson decay is lost, $W \rightarrow \tau\nu\gamma$ where the tau decays leptonically and any process which can produce
a lepton and an electron in the final state which fakes a photon.  The DZero analysis combines the electron and muon
channels and uses 162 pb$^{-1}$ of data for the electron channel and 82 pb$^{-1}$ for the muon channel.  A total of 146 events
is found in the electron channel with an estimated background of $87.1 \pm 7.5$ events, while 77 events are found in the
muon channel with an expected background of $37 \pm 10$ events.  This leads to a cross section times branching ratio of
$19.3 \pm 2.7(stat) \pm 6.1(syst) \pm 1.2(lumi)$ pb that compares well with the Baur NLO calculation of $16.4 \pm 0.4$ pb for
$W\gamma$ production with photon $E_T > 8$ GeV.
The CDF analysis finds 259 events after combining electron and muon channels in 202 pb$^{-1}$ of data.  The estimated
background is $75.12 \pm 15.01$ events and the cross section times branching ratio is 
$19.7 \pm 1.7(stat) \pm 2.0(syst) \pm 1.1(lumi)$ pb.  The Baur NLO prediction for the cross section with photon $E_T > 7$ GeV is
$19.3 \pm 1.3$ pb and is in good agreement with the CDF measurement.
Neither experiment observes an excess of high $E_T$ photons and both of the experiments are in the process of 
setting limits on anomalous couplings.

Similar analyses have been pursued for $Z\gamma$ production.  While the same initial and final state radiation contributions 
exist in $Z\gamma$ production as in $W\gamma$ production, the Standard Model forbids radiation of a photon directly from a
$Z$ boson, and no triple gauge coupling contribution is expected.  Thus, the triple gauge coupling contribution destructively
interferes in $W\gamma$ production but not in $Z\gamma$ production, and the relative background contribution from $Z$+jet production
is smaller than the relative contribution of $W$+jet production to $W\gamma$ events.  CDF has measured $Z\gamma$ production for
central photons with $E_T >$ 7 GeV that satisfy $\Delta R > 0.7$ in both the electronic and muonic $Z$ boson decay channels in
202 pb$^{-1}$ pb of data.  A total of 69 events is observed with an estimated background of $4.4 \pm 1.3$ events.  This yields
a cross section of $5.3 \pm 0.6(stat) \pm 0.4(syst) \pm 0.3(lumi)$ pb which compares well with the Baur NLO calculation
of $5.4 \pm 0.4$ pb.  Again, no excess of high energy photons is observed, and limits on anomalous couplings are forthcoming.

Pairs of $W$ bosons can also be produced in Tevatron collisions and constitute an important background to some Higgs searches
where the Higgs is expected to decay to $W$ boson pairs.  Analyses searching for H$\rightarrow W^+W^-$ are discussed elsewhere in
these proceedings \cite{nagy}.  Backgrounds to $WW$ production include
Drell-Yan, other diboson production, $Z \rightarrow \tau^+\tau^-$ where the taus decay leptonically, and top quark pair production.
Two analyses from CDF measure production of $W$ boson pairs.  The first analysis is a high purity
analysis that uses events with two high $E_T$ leptons with opposite charge, missing transverse energy greater than 25 GeV and no
high energy jets.  Events in this analysis are rejected if the dilepton mass is near the $Z$ boson mass and the missing 
transverse energy significance is low.  
A second, higher
efficiency analysis requires events to have one high transverse energy lepton and one isolated high transverse momentum track.
Both analyses use 184 pb$^{-1}$ of data, and the high purity analysis finds 17 events with an expected background of $4.8 \pm 0.7$
events.  This results in a cross section measurement of $14.2 \pm ^{5.6}_{4.9}(stat) \pm 1.6(syst) \pm 0.9(lumi)$ pb.  The
high efficiency analysis finds 39 events with a background of $15.1 \pm 0.9$ events and a cross section of
$19.4 \pm 5.1(stat) \pm 3.5(syst) \pm 1.2(lumi)$ pb.  Both of these cross sections are in agreement with the NLO calculation
by Ellis and Campbell, which predicts $12.5 \pm 0.8$ pb.
 
Finally, the production of $ZZ$ and $WZ$ boson pairs is searched for at CDF by looking for two to four leptons in electron
and muon channels.  
The analysis uses 194 pb$^{-1}$ and requires one lepton pair to be consistent with the mass of the
$Z$ boson. Four events are observed with a background estimated to be $5.1 \pm 0.7$ events, and the production cross section 
of $ZZ$ or $WZ$ is determined to be less than 13.8 pb at the 95\% confidence level.  A Standard Model calculation by Ellis and
Campbell predicts the cross section to be 5.2 pb.

\section{Electroweak Fits}

A large number of precision measurements of various parts of electroweak physics has been reported by experiments at
LEP, SLC and the Tevatron.  At the $Z$-pole, measurements of the $Z$ lineshape, polarized leptonic asymmetries, heavy flavor
asymmetries, heavy flavor branching fractions and hadronic charge asymmetry have been conducted by experiments at LEP and SLC.  
Additionally, the mass of the $W$ boson has been measured by experiments using the LEP-2 collider and the Tevatron, and the top quark
mass has been measured at the Tevatron.  In the Standard Model, each of these observables can be calculated in terms of
other electroweak parameters.  The mass of the Higgs and the top quark enter as radiative corrections of
order 1\%.  Programs such as ZFITTER and TOPAZ0 are used
\cite{lepewwg} to fit to the observables
and make predictions for variables such as the mass of the Higgs.  Important contributions to the uncertainty of the predicted
mass of the Higgs come from the uncertainties on the masses of the $W$ boson and top quark.  Run II of the Tevatron provides
a unique environment to measure both masses to precisions not before attainable.  

Measurements of the $W$ boson's mass at LEP-2 are accomplished by fully reconstructing $e^+e^- \rightarrow W^+W^- \rightarrow qqqq$ 
or $qq\ell\nu$ events.  The difference in the mass measured in the two different final states is $22 \pm 43$ MeV and the
mass measured by combining results from all LEP experiments is $80.412 \pm 0.042$ GeV.  The measurements at the Tevatron 
rely on fits to the tail of the transverse mass spectrum.  Using data from Run I of the Tevatron, combining the DZero and CDF
results gives a measured mass of $80.452 \pm 0.059$ GeV.
Significant contributions to uncertainties uncorrelated between the CDF and DZero experiments arise
from the number of $W$ bosons collected, the lepton energy scale and modeling of the hadronic recoil.  The increased
luminosity available in Run II along with upgrades to the detectors will reduce these uncertainties.  Additionally, the uncertainties
in the proton's parton distribution functions lead to significant errors correlated between the experiments.  These
uncertainties are expected to be reduced as information from the $W$ boson charge asymmetry and $Z$ boson rapidity distribution
is used to constrain the parton distribution functions.  Each experiment expects to achieve a
precision of $\sim0.040$ GeV on the measurement of the mass of the $W$ boson.  Combining the results from the two detectors
should result in an uncertainty of $\sim$0.025 GeV and would improve the current world average measurement of
$80.425 \pm 0.034$ GeV.

The mass of the top quark was measured in Run I and a recent update by DZero \cite{DZeroTop} has resulted in $\sim15\%$ smaller errors on that
measurement.  The details of the analysis are discussed elsewhere in these proceedings \cite{hocker} and result in a measurement of the top
quark mass of $180.1 \pm 5.3$ GeV.  Inclusion of this update in the combined Tevatron average gives a top quark mass
of $178.0 \pm 4.3$ GeV, which represents an increase of 3.7 GeV from the previous Tevatron average.  
A preliminary measurement of the top quark mass in Run II has been completed by CDF and is reported
elsewhere in these proceedings \cite{hocker} but is not yet included in the electroweak fits.  The expected accuracy of the measurement of the
top quark mass in Run II is 2.5 GeV.

Comparing the values for the masses of the $W$ boson and top quark obtained by the electroweak fits to direct measurements
is a test of the performance of the Standard Model.  The top
quark mass predicted by the fit is $178.5 \pm 9.7$ GeV, which is in good agreement with the Tevatron measurement.
The predicted value for the mass of the
$W$ boson of $80.386 \pm 0.023$ is also in good agreement with the direct measurements; however, the error on the direct
measurement is larger than the error of the prediction.  The precision of the Run II measurements of the $W$ boson's mass is expected to be
similar to the error of the prediction.

Fitting to the LEP, SLC and Tevatron data is also used to predict the mass of the Higgs.  The resulting most likely mass of
the Higgs particle is $114 \pm ^{62}_{42}$ GeV with a 95\% confidence level upper limit of 237 GeV.  
Improving the precision of the
measurement of the top quark mass with the inclusion of Run II data can be clearly seen to be
important because the relatively small increase
of 3.7 GeV in the measured top quark mass results in an increase of nearly 20 GeV in the most likely value for the mass
of the Higgs particle.  

\section{Conclusion}

A large number of electroweak measurements by CDF and DZero using Tevatron Run II data has been presented, and the measurements 
are in agreement with
the Standard Model.
Detector
understanding is increasing, as are the sizes of the data sets available.  These increases in precision are leading to preliminary $W$ boson
mass measurements with a combined Tevatron goal of 25 MeV accuracy.  The $W$ boson mass measurements along with improving top 
quark mass measurements are important elements of electroweak fits which predict the Higgs mass.  Using the most recent updates to the
Run I top quark mass measurement, the Higgs mass is constrained to be less than 237 GeV at the 95\% confidence level.

\section{Acknowledgements}

I would like to thank the organizers of PiC 2004 an interesting and informative series of lectures.  I would also like to thank my
colleagues from DZero, CDF and the LEP Electroweak Working group for their help in the preparation of this presentation.

\end{document}